\documentclass[slac_one]{revtex4}
\usepackage{graphicx}
\usepackage{fancyhdr}
\begin{document}

\title{Photoproduction at RHIC and the LHC} 

\author{Spencer R. Klein}
\affiliation{Nuclear Science Division, LBNL, Berkeley, CA, 94720 USA}

\begin{abstract}
The strong electromagnetic fields carried by relativistic highly charged ions make heavy-ion colliders attractive 
places to study photonuclear interactions and two-photon interactions.   
 At RHIC, three experiments have studied coherent photoproduction of $\rho^0$, 4$\pi$, $J/\psi$,
$e^+e^-$ pairs, and $e^+e^-$ pairs where the electron is bound to one of the incident nuclei.  These
results show that photoproduction studies are possible, and demonstrate some of the unique possibilities 
due to the symmetric final states and the ion targets.  
The LHC will reach  photon-nucleon energies many times higher than at HERA; these collisions can be used to measure
the gluon distributions in nuclei at very low Bjorken$-x$, where shadowing and gluon saturation may become important;
LHC $\gamma\gamma$ collisions may also be attractive places to search for some types of new physics. 
ATLAS, CMS and ALICE  are all planning to study photoproduction.  
After introducing the principles of photoproduction at hadron colliders, I will review recent results from RHIC 
on meson  and $e^+e^-$ production, and then discuss prospects for studies at the LHC.  

\end{abstract}

\maketitle

\thispagestyle{fancy}

\section{INTRODUCTION}

Photoproduction has traditionally been studied with photon beams at fixed target accelerators
and in $ep$ collisions at HERA.  However, photoproduction may also be studied at hadron colliders; the
photons come from the electromagnetic fields accompanying relativistic nuclei \cite{reviews}. 
The photon flux scales as the nuclear charge, $Z$, squared, so heavy nuclei generate intense fields.
Both photonuclear and 'two-photon' interaction can be studied.  These are known as `ultra-peripheral
collisions,' or UPCs. 

The maximum photon energy from a nucleus with radius $R_A$ and Lorentz boost 
$\gamma$ is $\gamma\hbar c/R_A$, or, for heavy ions about 3 GeV with RHIC and 100 GeV at the LHC.
This corresponds to $\gamma p$ center of mass energies of 25 and 800 GeV respectively. 
Light ions can reach considerably higher energies; the latter is four times higher than is available
at HERA.  Until
a new $ep$ collider, is built, the LHC will provide the highest energy photon-nucleon 
and two-photon collisions in the world. 

Besides the higher energies, heavy ion colliders have other advantages.
Their electromagnetic fields are intense enough so that three and four photon interactions
can be observed, so one can study multiple interactions involving a single ion pair. 
Relativistic heavy-ions are highly charged, so the resulting electromagnetic fields are very strong.
The common impact parameter leads to correlations in photon energy and linear polarization. 
So far, three and four photon reactions have been studied at RHIC.  Bound-free pair production (BFPP)
is another unique reaction. In it, an $e^+e^-$ pair is produced with the electron
bound to one of the incident nuclei.  This process produces a beam of single-electron ions; at the 
LHC this beam may carry enough power to quench superconducting magnets, limiting the achievable luminosity.

After reviewing current results, this writeup will discuss the 
future physics possibilities at the LHC.

\section{RESULTS FROM RHIC AND THE TEVATRON}

The first study of photoproduction at hadron colliders was a measurement of two photon production
of $\mu^+\mu^-$ pairs at the CERN Intersecting Storage Rings (ISR) $pp$ collider \cite{ISR}.  
In the 1980s, $e^+e^-$ production was studied at fixed target accelerators.  More recently, 
UPCs have been studied at RHIC at Brookhaven National Laboratory and the Fermilab Tevatron.  At
RHIC, the STAR and PHENIX experiments have studied a number of leptonic and hadronic final states, 
and a 3rd experimental collaboration has measured BFPP with copper beams \cite{bruce}.

The pioneering STAR studies of $\rho^0$ photoproduction and decay to $\pi^+\pi^-$
established the basic experimental 
parameters for UPCs.  For this, STAR used a trigger based on slats of scintillator 
surrounding a
central detector to select events with low muliplicity.  The slat hits were required to be
roughly back-to-back transversely, as expected from $\pi^+$ and $\pi^-$ produced in the
decay of a low $p_T$ $\rho^0$. 
The initial analysis selected events with $\rho^0$ $p_T < 150$ MeV/c, as is expected in coherent
photoproduction \cite{STARrho1}.  Like-sign pion pairs ($\pi^+\pi^+ + \pi^-\pi^-$) were a 
measure of the background.  

\begin{figure}[t]
\begin{minipage}{3 in}
\includegraphics[clip,scale=0.32]{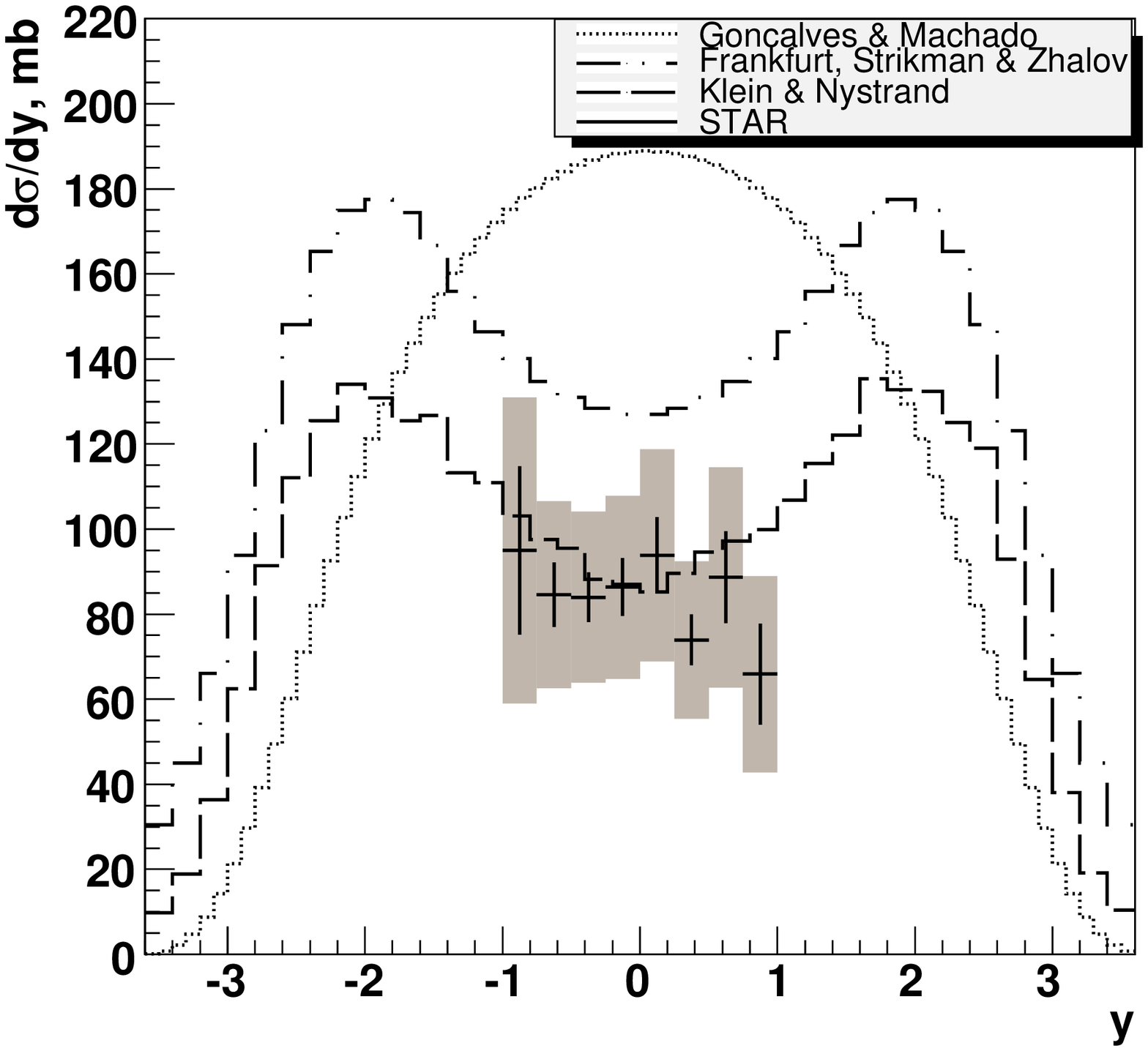}
\caption{
STAR measurement of $d\sigma/dy$ for $\rho^0$ photoproduction,
compared with three theoretical models \cite{STARrho2}.  The Goncalves and Machado calculation is based on a
saturation model, while the other two models are based on Glauber calculations.
\label{fig:STARrho}
}
\end{minipage}
\hskip 0.5 in
\begin{minipage}{3 in}
\includegraphics[clip,scale=0.38]{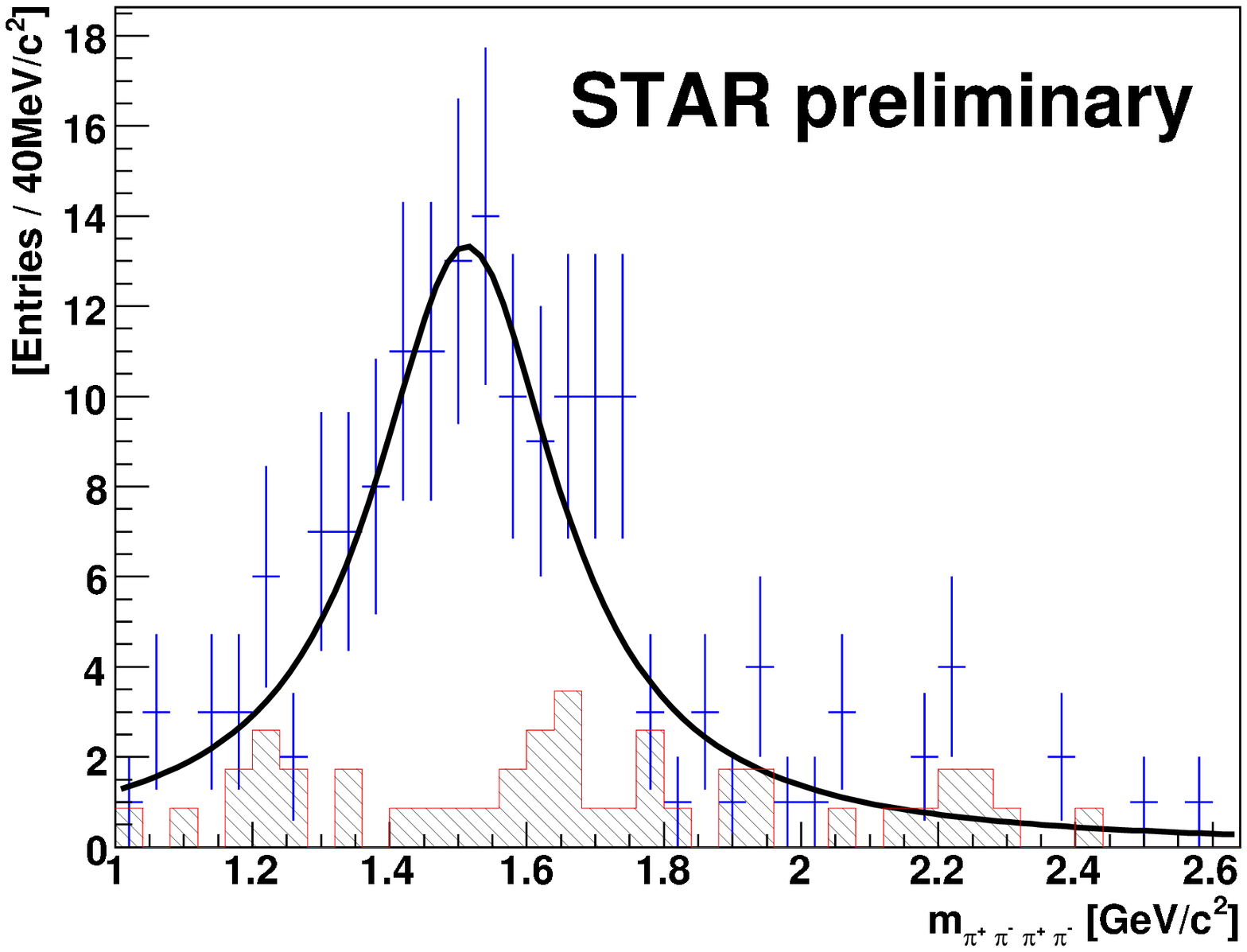}
\vskip .08 in
\caption{
Invariant mass for coherently photoproduced four-pion final states from the STAR 2004 run \cite{Boris}.
The points with error bars are net charge 0 ($\pi^+\pi^-\pi^+\pi^-$), while the shaded histogram is
a background estimate from  the net charge 2 ($\pi^+\pi^+\pi^+\pi^-$ + $\pi^+\pi^-\pi^-\pi^-$) data.
\label{fig:STAR4prong}
}
\end{minipage}
\end{figure}

STAR found that the coherent production cross-section agreed with theoretical calculations 
based on the Weiszacker-Williams  photon flux, convoluted with
photoproduction cross-sections determined by using a Glauber calculation \cite{Glauber}.  The
calculations used HERA data on $\gamma p\rightarrow \rho^0 p$ as input. 
Figure \ref{fig:STARrho} compares the $\rho^0$ rapidity distribution with three calculations, including
two based on Glauber models.   
STAR also observed direct $\pi^+\pi^-$ photoproduction (without the $\rho^0$ intermediary), through it's
interference with $\rho^0\rightarrow\pi^+\pi^-$.  
Later STAR studies of $\rho^0$ photoproduction expanded to include incoherent photoproduction
(where the photon scatters from a single nucleon, and the final state $\rho^0$ has higher $p_T$) \cite{STARrho2},
and also observed $\rho^0$ photoproduction in $dA$ collisions. 

STAR also studied $\rho^0$ photoproduction accompanied
by mutual Coulomb excitation, whereby the two nuclei are electromagnetically excited.  This reaction occurs
primarily by three-photon exchange - one for each nuclear excitation, plus a third to produce the $\rho^0$.
The two neutrons produced by the nuclear deexcitation make a simple and convenient
experimental trigger; these results have smaller systematic errors than the corresponding exclusive $\rho^0$ results.
The photons act independently, but, their common impact parameter introduces correlations \cite{factorize}.
The mutual Coulomb excitation  `tag' events with smaller average
impact parameters, and consequently, a harder photon spectrum \cite{baltzus}.

More recently, STAR has studied the photproduction of four-pion final states.  They are expected to
be produced primarily through excited $\rho$ states, most notably the $\rho^0(1450)$ and/or the $\rho^0(1700)$.
STAR observes a broad peak with a mass of roughly 1500 MeV/c$^2$, and a width of a few hundred MeV/c$^2$ \cite{Boris};
the yield is a small fraction of the $\rho^0$. Figure \ref{fig:STAR4prong} shows the $4\pi$ invariant mass
distribution; this distribution is compatible with earlier photoproduction studies. 

\begin{figure}[t]
\begin{minipage}{3 in}
\includegraphics[clip,scale=0.32]{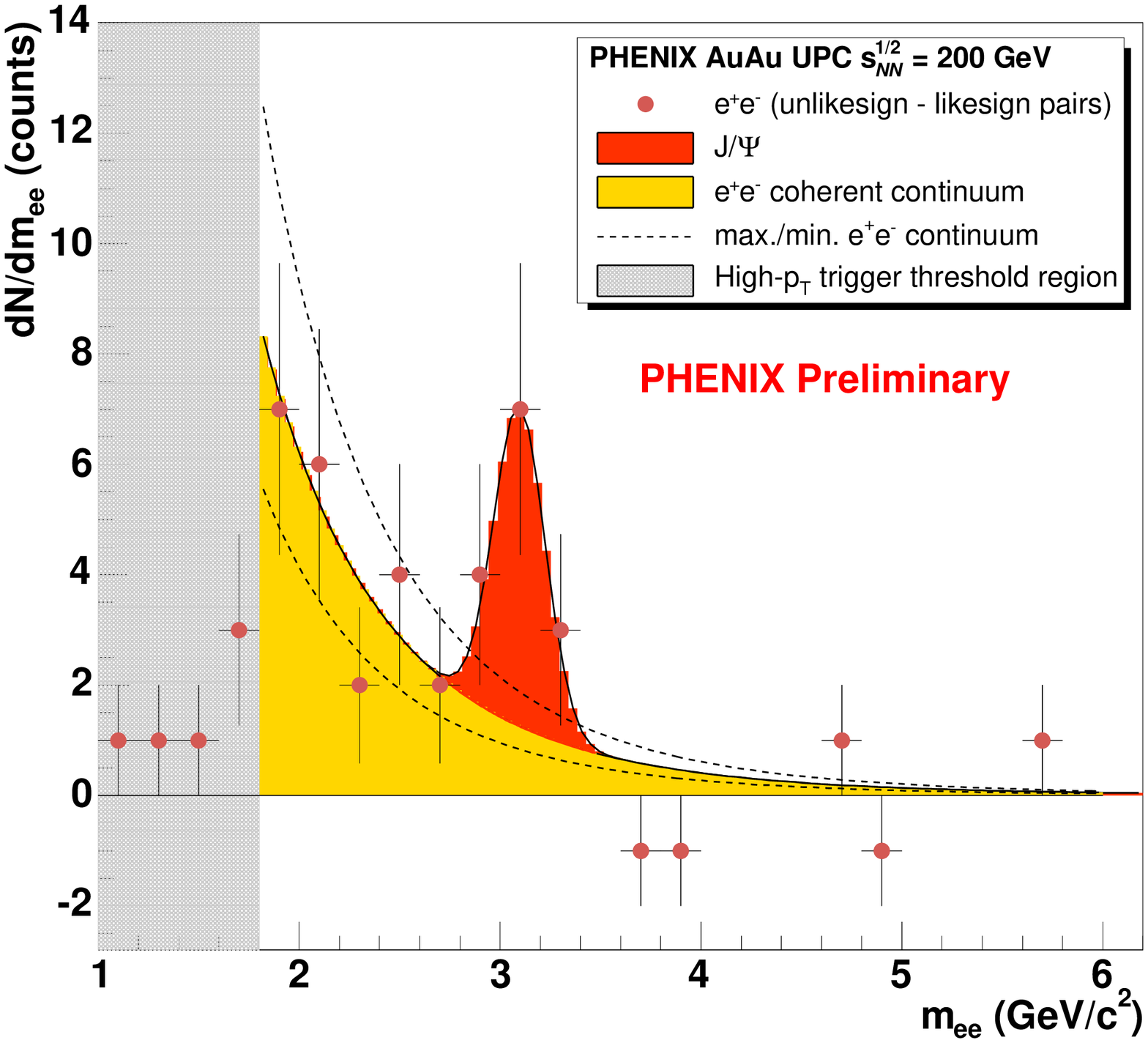}
\caption{
PHENIX measurements of coherent $e^+e^-$ production.  The yellow shading shows the
expectation from contiuum (electromagnetic) $e^+e^-$ production, while the red Gaussian 
shows the $J/\psi$ decay to $e^+e^-$  \cite{PHENIXJpsi}.  
\label{fig:PHENIXee}
}
\end{minipage}
\hskip 0.5 in
\begin{minipage}{3 in}
\includegraphics[clip,scale=0.35]{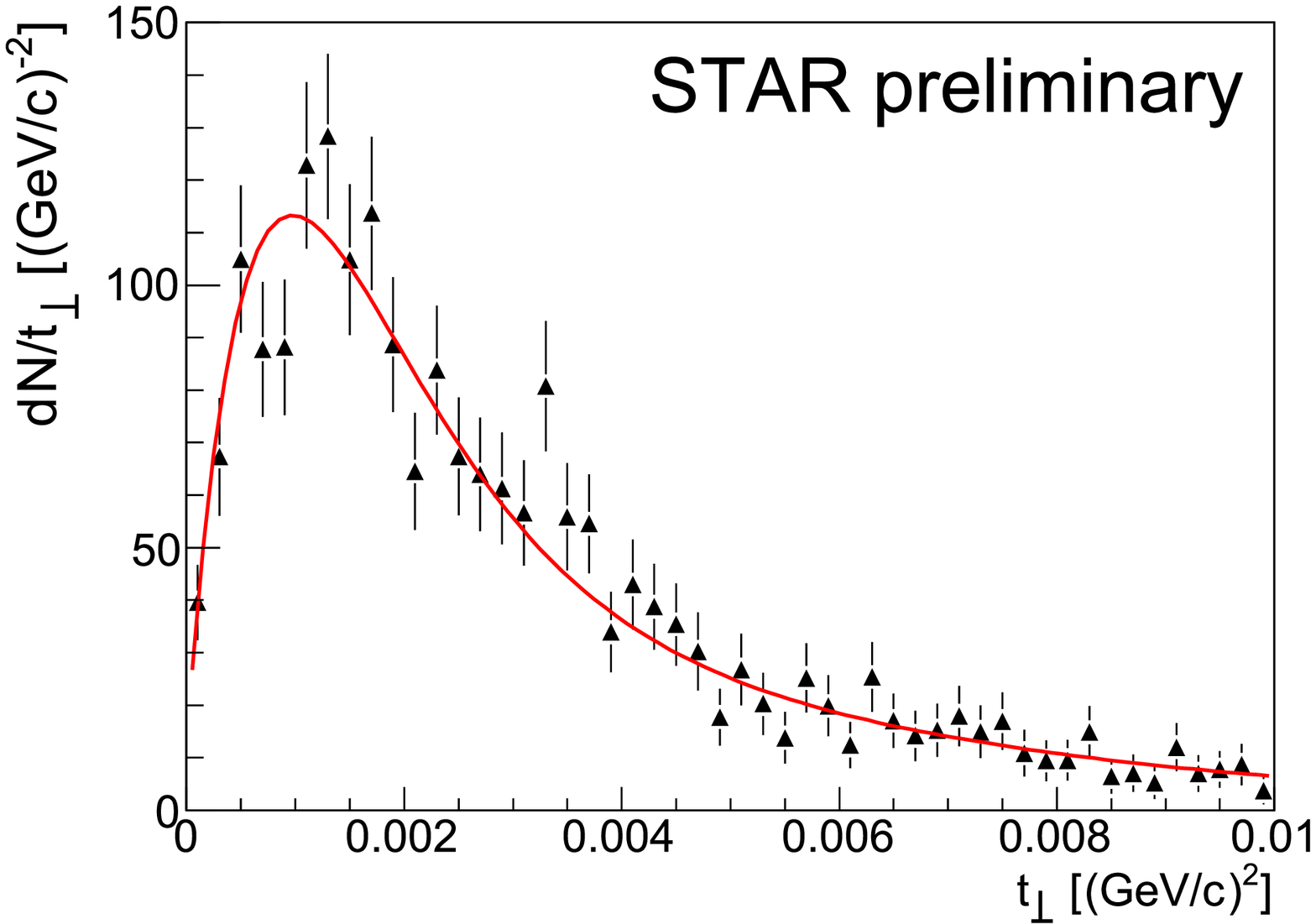}
\caption{
The $t_\perp = p_T^2$ spectrum for
$\rho^0$ photoproduction accompanied by mutual Coulomb excitation in STAR, for $0.1 < |y|<0.5$ \cite{QM2004}. 
\label{fig:STARinterfere}
}
\end{minipage}
\end{figure}

One unique aspect of UPC photoproduction is due to the initial state symmetry. 
Photoproduction can occur through two indistinguishable channels: either nucleus \#1
({\it e.g. from the blue beam})
can emit a photon which scatters from nucleus \#2 ({\it e.g. from the yellow beam}), or nucleus
\#2 emits a photon which scatters from nucleus \#1.  These two possibilities are
indistinguishable, and are related by a parity transformation.  Because vector mesons are
negative parity, the two amplitudes interfere destructively, and, at mid-rapidity, where the
two amplitudes $A$ are the same, the cross-section is \cite{interfere,BHT}
\begin{equation}
\sigma(b,y=0) = |A(b,y=0)|^2 \big[ 1-cos(\vec{p_T}\cdot\vec{b}) \big]
\end{equation}
where $\vec{b}$ is the impact parameter vector.  This interference suppresses production 
at low $p_T$.  Fig. \ref{fig:STARinterfere} shows the $t_\perp = p_T^2$ spectrum
for $\rho^0$ photoproduction with $0.1 < |y|<0.5$; the downturn for $t_\perp < 0.001$ GeV$^2$
is due to the interference \cite{QM2004}.

STAR has also studied purely electromagnetic production of $e^+e^-$ pairs \cite{STARee},
again acompanied by mutual Coulomb excitation.  Because of the low $p_T$ of the individual
leptons, the neutrons from the Coulomb excitation were necessary for triggering. 
Although this process is often described as `two-photon' production, a recent theoretical calculation
finds that higher order corrections are required to describe the data \cite{baltzee}.

PHENIX has focused on higher mass final states, using a calorimetric electron trigger.  Figure \ref{fig:PHENIXee} shows the
mass spectrum of coherently produced $e^+e^-$ final states \cite{PHENIXJpsi}. Electromagnetic $e^+e^-$ production
is seen at a rate consistent with that expected for `two-photon' production.  Above it, a
$J/\psi$ peak is observed, containing $10\pm 3$ events.  

CDF has recently published a 16-event mass spectrum for
exclusive $e^+e^-$  production in 1.8 TeV 
$\overline p p$ collisions at the Fermilab Tevatron\cite{CDFee}. A more recent CDF study of
$\mu^+\mu^-$ production found 334 events, including continuum $\mu^+\mu^-$, $J/\psi\rightarrow\mu^+\mu^-$
and $\psi'\rightarrow \mu^+\mu^-$ \cite{pinfold}. 
For protons, one background to
$J/\psi$ photoproduction is from double-Pomeron production of $\chi_c$, followed by 
$\chi_c\rightarrow J/\psi\gamma$, with the low energy photon
escaping the detector.  

Another important UPC reaction is bound-free pair production (BFPP)  where an $e^+e^-$ pair is produced 
with the electron bound to one of the incident ions, leaving a
single-electron atom \cite{bfpp}.  Since charge, but little momentum is transferred to the atom, BFPP produces a 
well collimated beam of ions, with  increased magnetic rigidity.  With lead beams at the LHC, the BFPP cross-section
is about 280 barns; the resulting beam of single-electron lead will strike the beampipe about 380 m downstream from the interaction region.  
At full LHC luminosity, this beam carries about 25 W of power, potentially enough to quench the
target superconducting dipole magnet \cite{beampipe}.
BFPP was recently measured with copper beams at RHIC.  Showers
from single-electron copper ions striking the RHIC beampipe were observed about 140 m downstream from the 
interaction point, at a rate consistent with the expected 0.2 b cross-section \cite{bruce}.

\section{PLANS FOR THE LHC}

The prospects for studying high-energy photoproduction with  UPCs at the LHC has attracted considerable
recent attention \cite{LHC}.  UPCs can be used to probe many aspects of physics. Here, we focus on three
topics.  

Considerable attention has been paid to measurements of the parton (particularly gluons) distribution 
functions of nuclei.  Several approaches are under consideration.
Experimentally, the simplest final state is heavy quarkonium, the $J/\psi$ or the
$\Upsilon.$  Photoproduction of a vector meson with mass $m_V$ occurs when a photon with energy $k$ interacts 
with a gluon with momentum fraction $x= m_V^2/4km_p$ and virtuality $Q^2 = (M_V/2)^2$; $k$ depends on the 
rapidity of the vector meson: $k = m_V/2 exp(\pm y)$. Of course, a second, soft gluon is required to conserve
momentum and color charge. 

The two-fold ambiguity in $k$ is due to the unknown photon 
direction; it disappears at $y=0$, and may be resolved elsewhere by comparing photoproduction with and 
without nuclear breakup.   With this, vector meson photoproduction can probe x values from $10^{-2}$ 
down to $10^{-5}$.  
Other channels of interest include heavy quark photoproduction (possibly 
including the top) \cite{heavyquarks} and studies of $\gamma$-jet and two-jet photoproduction \cite{Ramona}.

Another physics topics of interest is the study of the 'black disk' regime of QCD.  This occurs 
at high enough photon energies so that one 
probes low enough $x$ and $Q^2$ values so that one reaches the unitarity limit and
the target nucleus is essentially black.  In this regime, new phenomena appear.  The regime is best
illustrated by treating photons as $q\overline q$ dipoles, with varying sizes.  In this regime,
dipoles with smaller and smaller sizes interact.  These small sizes correspond to higher masses, so
high-mass diffraction is greatly enhanced.  

Finally, UPCs are sensitive to some types of new physics.  In particular,
two-photon production of the Higgs is of great interest.  Unfortunately, the cross-sections are small, but,
using protons beams, with forward detectors to tag the scattered protons, it may be possible to see a 
signal; although unlikely to be a discovery channel, this approach will be important in characterizing the Higgs.
Searches for other new phenomena, such as magnetic monopoles are also of great interest. 

The ALICE, ATLAS and CMS collaborations are all planning UPC programs.  These programs are largely focused on
photoproduction of $J/\psi$, $\psi'$ and $\Upsilon$, because of the relative experimental feasibility and the
physics interest.  For all three detectors, triggering is the key issue; dilepton decays of heavy quarks 
offer relatively simple triggering.  A secondary challenge, particularly for ATLAS, is being able to reconstruct 
charged particle tracks down to sufficiently low $p_T$.  There is also interest in ALICE to study $e^+e^-$ pair production using the inner silicon detector.
The cross-sections are large enough that a trigger is apparently not needed \cite{LHC}. 

\section{CONCLUSIONS}

Ultra-peripheral collisions have developed greatly since the first  studies of $e^+e^-$ and
$\mu^+\mu^-$ production at the ISR and fixed-target ion accelerators.  Work at RHIC has 
proved the feasibility of a variety of measurements and shown the value of the colliding beams
geometry for unique measurements.  

At the LHC, the three large experiments are planning a significant program, with measurements of
heavy-quarkonium production at the most advanced stage; this should lead to a rapid measurement of
the gluon distributions in nuclei at low$-x$ and moderate $Q^2$. Later studies will look at a variety of
other final states, including additioanl measurements of gluon structure functions and studies of parton saturation,
and searches for new physics.  

\begin{acknowledgments}

This work was supported by the Director, Office of Science, Office of Nuclear Science of the U.S. 
Department of Energy under Contract No. DE-AC02-05CH11231.

\end{acknowledgments}

\end{document}